\documentstyle[preprint,eqsecnum,aps]{revtex}
\begin{document}
\draft
\preprint{UW/PT-93-04}
\title{
Is Geodetic Precession of Massive Neutrinos the
solution to the Solar Neutrino Problem?}
\author{Stamatis Vokos\cite{doe}}
\address{
Department of Physics, FM-15\\
University of Washington,\\
Seattle, WA 98195}
\date{\today}
\maketitle

\begin{abstract}
We examine a recent suggestion by Milburn that slightly massive
electron neutrinos,
produced left-handed at the core of the sun, suffer
geodetic precession adequate to render them
right-handed (and therefore sterile) in sufficient numbers
to solve the solar neutrino problem. In that light, we perform a complete,
general-relativistic calculation of the geodetic
spin precession of an ultrarelativistic particle in the Schwartzschild metric.
We conclude that the effect is negligible, in disagreement with Milburn's
analysis.
\end{abstract}

\widetext
\section{Introduction}
Recently, Milburn \cite{milburn} suggested a solution to the solar
neutrino problem which does not require any assumptions
beyond the existence of a small, but non-zero, mass for
the electron neutrinos, consistent with
experimental limits ($m_\nu \leq 10\, {\rm eV}$). These
neutrinos are produced in the core of the sun with typical values for
$\gamma=E_\nu/m_\nu$ of the order of $10^4$--$10^6$. If
produced eccentrically, they will suffer a small bending
due to the gravitational pull of the sun.
Milburn argued that these neutrinos
undergo a {\em Thomas} precession of their spin
given by the same formula
relating the spin precession angle of charged particles in an
accelerator to their bending angle \cite{cooper},
\begin{equation}
\theta_{\rm p} = -\gamma \theta_{\rm b}\,,\label{one}
\end{equation}
where in the case at hand $\theta_{\rm b}$ is
the angle of gravitational bending. The smallness of
$\theta_{\rm b}$ is overcompensated by the magnitude
of $\gamma$, which for reasonable values  of the
parameters, results in an adequate repolarization of the
originally left-handed neutrinos, turning them into
right-handed, and hence sterile for the purposes of weak
interactions.

Despite its ingenious simplicity, this suggestion is
wrong, inasmuch as it is based on the aforementioned
formula.  Spin precession calculations have been carried out in the
special cases of
circular orbits or arbitrary orbits with small
velocities (in the sense of the PPN formalism) \cite{straumann}.
Since both special cases are clearly inapplicable
to our current application,
we perform, in this letter, the complete
general-relativistic calculation of the spin precession
of a particle in the Schwartzschild
geometry.
 We obtain,
\begin{equation}
\theta_{\rm p} =
-{1\over 2\gamma} \theta_{\rm b}\,,
\end{equation}
which demonstrates that for ultrarelativistic
particles,  $\theta_{\rm p} \ll \theta_{\rm b}$,
and therefore unable to provide us with a
satisfactory rate of left- to right-handed
neutrino conversion.

\section{The Calculation}

We consider a test particle approaching a
body of mass $M$.  Its original direction is
along $\phi = 0$, while its final direction is
along the asymptote $\phi = \pi + \theta_{\rm b}$.
The Schwartzschild metric is
\begin{equation}
ds^2 = B(r) dt^2- A(r) dr^2- r^2
d\theta^2-r^2\sin^2\theta d\phi^2\,,
\end{equation}
where
\begin{equation}
B(r)=A(r)^{-1} = 1-2m/r\,,
\end{equation}
with $m=GM/c^2$. In the remaining analysis we use units
with $G=1=c$. Instead of considering this coordinate frame, it is
useful to construct an orthonormal frame (vierbein),
$e^\alpha\,_\mu$, and its inverse $e_\alpha\,^\mu$, where
\begin{equation}
g_{\mu\nu}=e^\alpha\,_\mu e^\beta\,_\nu
\eta_{\alpha\beta}\,.
\end{equation}
This frame is given by
\begin{equation}
e^0\,_t=\sqrt{B},\quad
e^1\,_r=1/\sqrt{B},\quad
e^2\,_\theta = r,\quad
e^3\,_\phi = r\sin\theta\,.
\end{equation}
To separate orthonormal indices from
coordinate indices, we will use
letters from the beginning of the greek alphabet ($\alpha,\beta,
\gamma,\ldots)$ for the former, while the latter shall be denoted by
greek letters from the middle of the alphabet
($\mu,\nu,\ldots$). Then the components of any vector $V$ can be expressed
as
\begin{equation}
V^\mu=e_\alpha\,^\mu V^\alpha\,,\qquad
V^\alpha = e^\alpha\,_\mu V^\mu\,.
\end{equation}
In particular, the four-velocity of the particle can be expressed as
\begin{equation}
u^\mu =
(\dot{t},\dot{r},\dot{\theta},\dot{\phi})\,,\qquad
u^\alpha = (\sqrt{B} \dot{t},\dot{r}/\sqrt{B},
r\dot{\theta},r\sin\theta \dot{\phi})\,,
\end{equation}
where $\dot{\mbox{}}={d\over ds}$ is the
derivative with respect to the proper time of the
particle.
Covariant derivatives are given in terms of the
connection coefficients $\omega_\alpha\,^\beta\,_\gamma$.
The only non-vanishing such coefficients are
\begin{eqnarray}
\omega_0\,^0\,_1 & = &\omega_0\,^1\,_0={d\over
dr}\sqrt{B}\,,\\
\omega_2\,^1\,_2 & = &\omega_3\,^1\,_3=-\omega_2\,^2\,_1=
-\omega_3\,^3\,_1=-{1\over r} \sqrt{B}\,,\\
\omega_3\,^3\,_2 & = &-\omega_3\,^2\,_3=r^{-1} \cot\theta\,.
\end{eqnarray}
The spin vector of the particle will in general be
Fermi-Walker transported along an arbitrary path $x^\beta(s)$ under
external non-gravitational forces \cite{straumann},
\begin{equation}
u^\alpha\nabla_\alpha S^\beta = - S^\gamma a_\gamma
u^\beta\,,
\end{equation}
where $a^\beta={Du^\beta\over ds}$ is the four-acceleration.
However, in the absence of non-gravitational forces
the four-acceleration is zero and the particle will move
along a geodesic, while its spin will be
parallel-transported,
\begin{equation}
u^\alpha\nabla_\alpha S^\beta = 0\,.\label{nabs}
\end{equation}
Thomas precession is a purely kinematical effect which
comes about when there is a non-zero four-acceleration,
for instance in an electron moving around a nucleus or in an
accelerator. For a particle moving along a
geodesic, the Thomas precession is identically zero.
Therefore, no direct analogy with special relativistic
electrodynamics can be drawn and the reasoning leading to (\ref{one})
 is incomplete.

However, the spin will still experience geodetic
precession. Eq.\ (\ref{nabs}) above becomes in components,
\begin{equation}
{dS^\alpha\over ds}+\omega_\beta\,^\alpha\,_\gamma u^\beta
S^\gamma = 0\,.
\end{equation}
Explicitly,
\begin{eqnarray}
{dS^0\over ds} & = &-(\sqrt{B})'u^0 S^1\,,\\
{dS^1\over ds} & = &-(\sqrt{B})'u^0 S^0 +{\sqrt{B}\over r} u^2
S^2 + {\sqrt{B}\over r}u^3 S^3\,,\label{Sdot}\\
{dS^2\over ds} & = &-{\sqrt{B}\over r} u^2 S^1 +{\cot
\theta\over r}u^3 S^3\,,\\
{dS^3\over ds} & = &-{\sqrt{B}\over r}u^3 S^1-{\cot
\theta\over r}u^3 S^2\,.
\end{eqnarray}
For motion in the equatorial plane, $\theta=\pi/2$ and
$u^2=0$. Furthermore the spin is a spacelike
vector, which in the rest frame of the particle has
vanishing zeroth component. Given that the four-velocity
is a timelike vector with vanishing spatial components in
the rest-frame of the particle implies that the spin is orthogonal to
the four-velocity.
This allows us to express $S^0$ as a linear combination
of the spatial components of $S$, namely
\begin{equation}
S^0 = {1\over u^0} (u^1S^1+u^3S^3)\,.
\end{equation}
Substituting into Eq.\ (\ref{Sdot}) gives three independent equations
\begin{eqnarray}
{dS^1\over ds} & = &-(\sqrt{B})'u^1 S^1 + \left({\sqrt{B}\over r}
- (\sqrt{B})'\right)u^3 S^3\,,\nonumber\\
{dS^2\over ds} & = &0\,,\label{sdot}\\
{dS^3\over ds} & = & -{\sqrt{B}\over r}u^3 S^1\,.\nonumber
\end{eqnarray}

We can simplify further the equations of motion of
the spin components, in view of the non-uniqueness of this
orthonormal basis $\{e^\alpha\,_\mu\}$.
We can use the freedom of choosing
an appropriate basis, in which the spin precession is
evident. This is achieved by going to the rest frame of
the particle. The basis corresponding to the rest-frame
$\{\bar{e}^\alpha\,_\mu\}$ is constructed through a
local Lorentz transformation with the following
properties: (a) the resulting zeroth component of the
spin vanishes and (b) the four velocity takes on the form
$(1,\vec{0})$.
Explicitly,
\begin{equation}
\bar{e}_\beta=e_\alpha L^\alpha\,_\beta\,,
\end{equation}
with
\begin{equation}
u_\alpha L^\alpha\,_i=0\,,\qquad u_\alpha L^\alpha\,_0=1\,,
\end{equation}
Then, $\bar{e}_i$ ($i=1,2,3$) are orthogonal to the
four-velocity, and we are guaranteed to get $\bar{S}^0=0$
when we resolve $S$ along the new basis,
\begin{equation}
S=S^\alpha e_\alpha = \bar{S}^i \bar{e}_i\,.
\end{equation}
To construct the requisite Lorentz transformation we
note that the vanishing of the $\alpha=0$ component of
\begin{equation}
\bar{S}_\alpha =S_\gamma L^\gamma \,_\alpha\,,
\end{equation}
together with $S\cdot u =0$, gives
\begin{equation}
{L^i\,_0 \over L^0\,_0}={u^i\over u^0}\,.
\end{equation}
Keeping in mind that a Lorentz transformation matrix
satisfies
\begin{equation}
\eta_{\alpha\delta} L^\alpha\,_\beta  L^\delta\,_\gamma=
\eta_{\beta\gamma}\,,
\end{equation}
we obtain
\begin{equation}
L^\alpha\,_0=u^\alpha\,.
\end{equation}
One can similarly obtain the remaining entries to arrive
at
\begin{eqnarray}
L^\alpha\,_\beta = \left(\begin{array}{cccc}
u^0 & u^1 & 0 & u^3\\
u^1 & 1+ {(u^1)^2\over 1+u^0} & 0 & {u^1u^3\over 1+u^0}\\
0 & 0 & 1 & 0\\
u^3 & {u^1u^3\over 1+u^0} & 0 & 1+ {(u^3)^2\over 1+u^0}
\end{array}\right)\,.
\end{eqnarray}
The equations of motion for the spin components  in the
rest frame can be inferred from (\ref{sdot}) with the aid of
the equation for the
vanishing of the four-acceleration,
\begin{equation}
{du^\alpha\over ds} + \omega_\beta\,^\alpha\,_\gamma
u^\beta u^\gamma =0\,.
\end{equation}
We obtain
\begin{eqnarray}
{d\bar{S}^1\over ds} &= &\bar{S}^3 {(r-2m+(r+3m)u^0)u^3\over
r^2\sqrt{B}(1+u^0)}\,,\nonumber\\
{d\bar{S}^2\over ds} &= &0\,,\label{sbardot}\\
{d\bar{S}^3\over ds} &= &-\bar{S}^1 {(r-2m+(r+3m)u^0)u^3\over
r^2\sqrt{B}(1+u^0)}\,.\nonumber
\end{eqnarray}
Next we replace the derivatives with respect to proper time
with derivatives with respect to coordinate time.
Recalling that
\begin{equation}
u^0=\sqrt{B} \dot{t}\,,\qquad
u^3= r\dot{\phi}\,,
\end{equation}
Eq.\ (\ref{sbardot}) obtains
\begin{equation}
{d\vec{S}\over dt}= \vec{\Omega} \times \vec{S}\,,
\end{equation}
where we dropped the bars for notational simplicity, and $\Omega^1=0=
\Omega^3$, while
\begin{equation}
\Omega^2\equiv\Omega=\left({r-3m\over r}+{m\over r(1+u^0)}\right){1\over
\sqrt{1-{2m\over r}}}{d\phi\over dt}\,.
\end{equation}
How can we turn the knowledge of $\vec{\Omega}$ into an
expression for the geodetic precession angle over the whole
trajectory of the particle? First, we recall that the
Schwartzschild geometry admits two Killing vectors, namely
${\partial\over \partial t}$ and ${\partial\over \partial
\phi}$. These two isometries correspond to two constants
of the motion,
\begin{equation}
E\equiv (1-{2m\over r}) \dot{t}\,,\quad
L\equiv r^2 \dot{\phi}\,.
\end{equation}
For a particle coming in from infinity,
\begin{equation}
E=\dot{t}|_{r\to\infty}\,.
\end{equation}
Asymptotically, the Schwartzschild metric becomes
Minkowskian, $ds^2=dt^2-d\vec{x}^2$, hence
\begin{equation}
E=\left.(1-{d\vec{x}^2\over dt^2})^{-{1\over 2}}\right|_{r\to\infty}
\equiv \gamma\,.
\end{equation}
We therefore obtain,
\begin{equation}
u^0={\gamma\over \sqrt{1-{2m\over r}}}\,.\label{u0}
\end{equation}
Denoting
\begin{eqnarray}
S_\pm &=& S^1\pm i S^3\,,\\
e(\phi)&=& \Omega/ {d\phi\over dt}\,,
\end{eqnarray}
the spin precession equations may be recast in the form
\begin{equation}
{dS_\pm \over d\phi}=\mp i e(\phi) S_\pm\,,
\end{equation}
with solution
\begin{equation}
S_\pm(\phi_{\rm f})=S_\pm(\phi_{\rm i}) e^{\mp i\int_{\phi_{\rm
i}}^{\phi_{\rm f}} d\phi\,e(\phi)}\,.
\end{equation}
In the absence of gravity $e(\phi)=1$, and if $\phi_{\rm i}=0$, then
$\phi_{\rm f}=\pi$, whereas in the presence of gravity $e(\phi)\neq 1$
and $\phi_{\rm f}=\pi+\theta_{\rm b}$. The total geodetic
deviation is
\begin{eqnarray}
\Delta S_\pm & = & S_\pm(\pi+\theta_{\rm b})-S_\pm^{(0)}({\rm
\pi})\nonumber \\
& = & S_\pm(\pi+\theta_{\rm b})+S_\pm (0)\,.\label{spm}
\end{eqnarray}
An ultrarelativistic particle follows approximately the same geodesic
as light, which to first order in $m/b$ is given by
\begin{equation}
u\equiv
{1\over r}={1\over b} \sin\phi + {3 m\over 2 b^2}(1+{1\over 3}\cos 2\phi)\,,
\label{u}
\end{equation}
where $b$ is the impact parameter. With the aid of Eqs.\ (\ref{u0})
and (\ref{u}),
\begin{eqnarray}
e(\phi)& = &{1-3mu\over \sqrt{1-2mu}}+ {mu\over \gamma+\sqrt{1-2mu}}\,,\\
& = & 1- mu(2-{1\over 1+\gamma}) + \cdots\,.
\end{eqnarray}
Substituting into Eq.\ (\ref{spm}) gives
\begin{eqnarray}
{\Delta S_\pm\over S_\pm(0)}&=&e^{\mp i (\pi + {1\over 2}\theta_{\rm b}/
\gamma)}+1\\
&=& \pm i {\theta_{\rm b}\over 2\gamma}+\cdots
\end{eqnarray}
This clearly agrees with the expectation that for light ($\gamma\to\infty$),
there is no excess spin precession.

\section{Discussion}

We have shown that geodetic precession of massive, initially left-handed,
neutrinos moving along geodesics in the Schwartzschild geometry,
cannot produce enough sterile neutrinos to solve the solar neutrino
puzzle. In fact, the answer is $\cal{O}(1/\gamma)=\cal{O}(m/E)$,
and conforms with conventional field-theoretic experience, in which
helicity flips vanish in the massless fermion (or ultrarelativistic) limit.
However, one may argue that our analysis is steeped in the spirit of
general relativity
and may rely on some special features which may not be shared by some other
interaction (i.e.\ one which does not satisfy the principle of equivalence).
In such a case, it could be imagined that a small deflection due to a
non-gravitational ``fifth force'' interaction could be amplified to
yield the requisite sterilizing effect suggested by Milburn.
But how can an interaction produce counter-intuitive $\cal{O}(E/m)$
helicity-flip probabilities? Milburn's argument suggests that
kinematics, not dynamics, can yield this effect. After all, in the
case of electrons deflected at SLAC \cite{cooper}, the vector couplings
of the QED lagrangian do not even connect left- and right-handed
spinors. But if such kinematical effects held for other interactions,
why would gravity (which, in many respects, can be modelled by
graviton exchange amplitudes) yield such a dramatically different
answer? It is, therefore, worth asking whether our result carries over
to other interactions, within the scope of Milburn's suggestion.

\acknowledgments
I am indebted to Cosmas Zachos for bringing this issue to my attention
and for useful conversations. I am also grateful
to David Boulware and Lowell Brown for several invaluable suggestions.

\end{document}